\documentclass[a4paper]{jpconf}

\usepackage{etex}
\usepackage{
  alltt,
  amsfonts,
  amsmath,
  amssymb,
  booktabs,
  caption,
  graphicx,
  hyperref,
  mathtools, 
  lineno, 
  pgfpages,
  pgfplots,
  siunitx,
  todonotes,
  tikz,
  url,
  xspace
}

\usepackage[percent]{overpic}  
\usepackage[absolute,overlay]{textpos}
\usepackage[utf8]{inputenc}

\usepackage[normalem]{ulem}  

\usepackage[square,sort&compress,numbers]{natbib}
\definecolor{mLightBrown}{HTML}{EB811B}
\definecolor{mRed}{HTML}{e41a1c}
\definecolor{mBlue}{HTML}{377eb8}
\definecolor{mGreen}{HTML}{4daf4a}
\definecolor{mPurple}{HTML}{984ea3}
\definecolor{mOrange}{HTML}{ff7f00}

\newcommand{\ie}{i.e.,\xspace}

\newcommand{\deta}{\ensuremath{\Delta\eta}\xspace}
\newcommand{\detagap}{$|\deta|$-gap\xspace}

\newcommand{\mean}[1]{\ensuremath{\left \langle#1 \right \rangle}\xspace}

\newcommand{\snn}{\ensuremath{\sqrt{s_{\text{NN}}}}\xspace}

\newcommand{\hvnn}{\ensuremath{V_{n\Delta}}\xspace}
\newcommand{\hvnnee}{\ensuremath{\hvnn(\eta_a, \eta_b)}\xspace}
\newcommand{\hvssee}{\ensuremath{V_{2\Delta}(\eta_a, \eta_b)}\xspace}

\newcommand{\eeplane}{($\eta_a, \eta_b$)-plane\xspace}

\newcommand{\demin}{\ensuremath{\Delta\eta_{\text{min}}}\xspace}

\tikzset{
  every overlay node/.style={
    draw=black,fill=white,rounded corners,anchor=north west,
  },
}
\usetikzlibrary{arrows, shapes, backgrounds, snakes, quotes, angles, calc, patterns, math}
\tikzstyle{every picture}+=[remember picture]
\tikzstyle{na} = [baseline=-.5ex]

\begin{document}
\title{Factorization of two-particle distributions in AMPT simulations of Pb--Pb collisions at $\mathbf{\snn} $ = 5.02 TeV}
\author{Christian Bourjau (for the ALICE Collaboration)}
\address{Niels Bohr Institute, University of Copenhagen\\Blegdamsvej 17, 2100 Copenhagen, Denmark}

\begin{abstract}
  The flow ansatz states that the single-particle distribution of a given event can be described in terms of the complex flow coefficients $V_n$.
  Multi-particle distributions can therefore be expressed as products of these single-particle coefficients; a property commonly referred to as factorization.
  The amplitudes and phases of the coefficients fluctuate from event to event, possibly breaking the factorization assumption for  event-sample averaged multi-particle distributions.
  Furthermore, non-flow effects such as di-jets may also break the factorization assumption.
  The factorization breaking with respect to pseudorapidity $\eta$ provides insights into the fluctuations of the initial conditions of heavy ion collisions and can simultaneously be used to identify regions of the phase space which exhibit non-flow effects.
  These proceedings present a method to perform a factorization of the two-particle Fourier coefficients \hvnnee which is largely independent of detector effects.
  AMPT model calculations of Pb--Pb collisions at $\snn = \SI{5.02}{TeV}$ are used to identify the smallest \detagap necessary for the factorization assumption to hold.
  Furthermore, a possible \deta-dependent decorrelation effect in the simulated data is quantified using the empirical parameter $F_2^\eta$.
  The decorrelation effect observed in the AMPT calculations is compared to results by the CMS collaboration for Pb--Pb collisions at $\snn = \SI{2.76}{TeV}$.
\end{abstract}

\section{Introduction}
The Fourier coefficients of an event-sample averaged two-particle distribution are commonly described as
\begin{align}
  \hvnnee
  &= \mean{V_n(\eta_a) V_n^*(\eta_b)}, \\
  \label{eq:average-written-out}
  &= \mean{v_n(\eta_a) v_n(\eta_b) e^{in(\psi_n(\eta_a) - \psi_n(\eta_b))}},
\end{align}
where $v_n$ are the flow coefficients and $\psi_n$ are the symmetry planes.
Either of these two quantities may fluctuate from event to even due to varying initial conditions, thereby breaking the factorization of the sample average even for simulations of ideal hydrodynamics~\cite{Gardim_2013}.
By studying the factorization behavior of \hvnnee one can therefore infer the properties of such fluctuations.
Flow related analyses commonly assume that non-flow contributions decrease with an increasing $\eta$-separation of the particles.
In order to minimize the impact of non-flow effects on the measurement a minimal longitudinal separation between particles, referred to as \detagap, is therefore often applied.
Under the assumption that non-flow effects do not factorize identically to anisotropic flow it is possible to identify regions of the phase space where non-flow effects become negligible~\cite{1110.4809,1002.0534}.

Whether Eq.~\eqref{eq:average-written-out} may be written in a factorized form depends on the correlations between the four quantities $v_n(\eta_a)$, $v_n(\eta_b)$, $\psi_n(\eta_a)$, and $\psi_n(\eta_b)$.
These proceedings focus on the effect of symmetry plane decorrelation effects.
The phases at $\eta_a$ and $\eta_b$ are commonly assumed to be correlated with each other through a common symmetry plane angle $\Psi_n$, but may fluctuate from event to event.
The fluctuations are equally likely to occur in either direction which ensures that \hvnnee is a real quantity.
The observed average is attenuated due to these fluctuations which are therefore also referred to as \emph{decorrelation} effects.
If the fluctuations at $\psi_n(\eta_a)$ and $\psi_n(\eta_b)$ in Eq.~\eqref{eq:average-written-out} exhibit a dependence on $\deta = \eta_a - \eta_b$ it may cause a factorization breaking of \hvnnee.

\section{Observable definition}
The results presented here are exclusively based on the Monte-Carlo (MC) simulations and would therefore not require considering detector effects on the observables.
However, in order to present a generally applicable method, the analysis presented here is constructed around an observable which is largely independent of uncorrelated detector deficiencies.

At its core, this analysis is based on the single- and two-particle distributions averaged over many events.
The single-particle distribution $\hat{\rho}_1$ is given by
\begin{equation}
  \label{eq:rho1}
  \hat{\rho}_1(\eta, \varphi) = \left \langle\frac{d^2N}{d\eta d\varphi} \right\rangle
\end{equation}
where $N$ is the number of observed charged particles and $\varphi$ is the azimuthal coordinate.
In experimental measurements any azimuthal anisotropies in $\hat{\rho}_1$ are exclusively caused by detector effects.

The distribution of particle-pairs $\hat{\rho}_2$ is given by
\begin{equation}
  \label{eq:pair-density}
  \hat{\rho}_2(\eta_a, \eta_b, \varphi_a, \varphi_b) = \mean{\frac{d^4N_{\text{pairs}}}{d\eta_a d\eta_ d\varphi_a d\varphi_b}}
\end{equation}
where $N_{\rm pairs}$ denotes the number of particle pairs observed at $\eta_a, \eta_b, \varphi_a, \varphi_b$.

The two distributions $\hat{\rho}_1$ and $\hat{\rho}_2$ can be used to construct the \emph{normalized two-particle density} $r_2$ which is largely independent of uncorrelated detector effects~\cite{Vechernin201521,1311.3915}.
\begin{equation}
  \label{eq:r2}
  r_2(\eta_a, \eta_b, \varphi_a, \varphi_b) = \frac{\hat{\rho}_2(\eta_a, \eta_b, \varphi_a, \varphi_b)}{\hat{\rho}_1(\eta_a, \varphi_a) \hat{\rho}_1(\eta_b, \varphi_b)}
\end{equation}
The non-zero two-particle Fourier coefficients can then be computed by
\begin{equation}
  \label{eq:24}
  \hvnnee = {\left ( \frac{1}{2\pi} \right )}^2\int_0^{2\pi} \int_0^{2\pi} r_2(\eta_a, \eta_b, \varphi_a, \varphi_b) e^{-in\varphi_a} e^{in\varphi_b} d\varphi_a d\varphi_b
\end{equation}

\section{Factorization}
\label{sec:factorization-def}
The functional form of \hvnnee in the \eeplane is assessed with two different models.
Both models focused on the flow ansatz \ie{} individual events can be described in terms of single particle distributions.
Neither model attempts to describe any \emph{non-flow} processes such as di-jets or weak decays.

\subsection{Purely factorizing model (Model A)}
\label{sec:purely-fact-model}
This model assumes that the averaged two-particle coefficients may be described by a product of two identical functions $\hat{v}_n^A$ 
\begin{align}
  \hvnn(\eta_a, \eta_b) &= \mean{V_n(\eta_a) {V_n^*(\eta_b)}} \\
                        &= \mean{v_n(\eta_a) v_n(\eta_b) e^{in(\psi_n(\eta_a) - \psi_n(\eta_b))}} \\
  \label{eq:factorization-def}
                        &= \hat{v}_n^A(\eta_a) \hat{v}_n^A(\eta_b)
\end{align}
If $\psi_n(\eta_a)$ is always equal to $\psi_n(\eta_b)$ within each event and if the fluctuations of $v_n$ are uncorrelated along $\eta$, Eq.~\eqref{eq:factorization-def} holds and $\hat{v}_n^A(\eta)$ is the mean value of the event-by-event flow coefficients $v_n(\eta)$.
The degree to which the measured $\hvnnee$ is compatible with Model A provides a limit to the size of factorization-breaking fluctuations of the flow coefficients, event-plane decorrelations and non-flow effects.

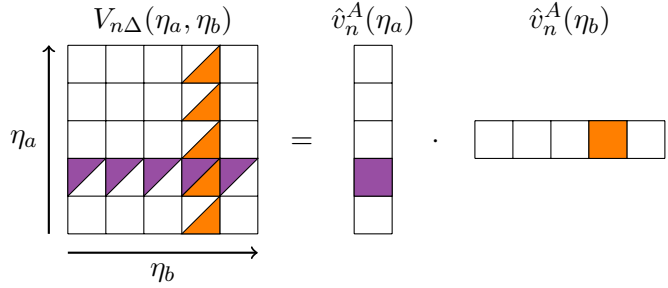
\begin{figure}
  \centering
  \[
    \begin{tikzpicture}[baseline=6.9ex, scale=0.5]
      \draw(0,0) grid (5,5);

      \foreach \x in {0,1,...,4} {
        \draw[fill=mPurple, xshift=\x cm, yshift=1cm] (0, 0) -- (1,1) -- (0, 1) -- cycle;
      }
      \foreach \y in {0,1,...,4} {
        \draw[fill=mOrange, xshift=3cm, yshift=\y cm] (0, 0) -- (1, 1) -- (1, 0) -- cycle;      
      }
      
      \draw[->, thick, xshift=-0.5cm] (0,0) -- node[left] {$\eta_a$} (0, 5);
      \draw[->, thick, yshift=-0.5cm] (0,0) -- node[below] {$\eta_b$} (5, 0);
      \node[above] at (2.5, 5) {$\hvnnee$};
    \end{tikzpicture}
    \hspace{0.8em}
    =
    \begin{tikzpicture}[baseline=6.9ex, scale=0.5]
      \draw (0,0) grid (1,5);
      \draw[fill=mPurple, yshift=1cm] (0, 0) rectangle (1,1);
      \node[above] at (0.5, 5) {$\hat{v}_{n}^A(\eta_a)$};
    \end{tikzpicture}
    \cdot
    \begin{tikzpicture}[baseline=6.9ex, scale=0.5]
      \draw[yshift=2cm] (0,0) grid (5,1);
      \draw[fill=mOrange, xshift=3cm, yshift=2cm] (0, 0) rectangle (1,1);
      \node[above] at (2.5, 5) {$\hat{v}_{n}^A(\eta_b)$};
      \node at (-0.5,0) {};
      \node at (5.5,0) {};
    \end{tikzpicture}
  \]
  
  \caption[Schematic representation of Model A]{Schematic representation of Model A. Each element of $\hat{v}_n(\eta)$ affects several elements of \hvnnee.}
  \label{fig:schema-model-a}
\end{figure}

The flow coefficients $\hat{v}_n^A(\eta)$ are fit to the observed \hvnnee.
The latter is computed as a histogram of finite bin size in $\eta_a$ and $\eta_b$.
Eq.~\eqref{eq:factorization-def} can thus be seen as a non-linear equation system
\begin{equation}
  \label{eq:equation-sys-mod-a}
  V_{n\Delta}(\eta_a^i, \eta_b^j) = \hat{v}_n^A(\eta_a^i) \hat{v}_n^A(\eta_b^j)
\end{equation}
where $i$ and $j$ are the bin-indices along $\eta_a$ and $\eta_b$ respectively.
Graphically, Eq.~\eqref{eq:equation-sys-mod-a} can be represented as shown in Fig.~\ref{fig:schema-model-a} where \hvnnee is a two-dimensional matrix and $v_n^A(\eta)$ a one-dimensional vector. 
Solving Eq.~\eqref{eq:equation-sys-mod-a} for all points in \hvnnee yields the ``vector'' $v_n^A(\eta)$ which best describes the observed data.
Each value in $\hat{v}_n^A$ affects several points in the \eeplane.

\subsection{Long-range decorrelating model (Model B)}
\label{sec:long-range-decorr}
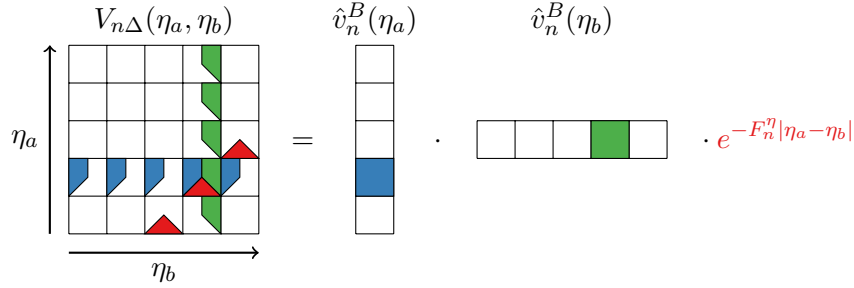
\begin{figure}
  \centering
  \[
    \begin{tikzpicture}[baseline=6.9ex, scale=0.5]
      \draw(0,0) grid (5,5);

      \foreach \x in {0,1,...,4} {
        \draw[fill=mBlue, xshift=\x cm, yshift=1cm] (0, 0) -- (0.5, 0.5) -- (0.5,1) -- (0, 1) -- cycle;
      }
      \foreach \y in {0,1,...,4} {
        \draw[fill=mGreen, xshift=3cm, yshift=\y cm] (1, 0) -- (0.5,0.5) -- (0.5, 1) -- (1, 1) -- cycle;      
      }
      \draw[fill=mRed, xshift=3 cm, yshift=1 cm] (0, 0) -- (0.5,0.5) -- (1, 0) -- cycle;
      \draw[fill=mRed, xshift=2 cm, yshift=0 cm] (0, 0) -- (0.5,0.5) -- (1, 0) -- cycle;
      \draw[fill=mRed, xshift=4 cm, yshift=2 cm] (0, 0) -- (0.5,0.5) -- (1, 0) -- cycle;
      
      \draw[->, thick, xshift=-0.5cm] (0,0) -- node[left] {$\eta_a$} (0, 5);
      \draw[->, thick, yshift=-0.5cm] (0,0) -- node[below] {$\eta_b$} (5, 0);
      \node[above] at (2.5, 5) {$\hvnnee$};
    \end{tikzpicture}
    \hspace{0.8em}
    =
    \begin{tikzpicture}[baseline=6.9ex, scale=0.5]
      \draw (0,0) grid (1,5);
      \draw[fill=mBlue, yshift=1cm] (0, 0) rectangle (1,1);
      \node[above] at (0.5, 5) {$\hat{v}_{n}^B(\eta_a)$};
    \end{tikzpicture}
    \cdot
    \begin{tikzpicture}[baseline=6.9ex, scale=0.5]
      \draw[yshift=2cm] (0,0) grid (5,1);
      \draw[fill=mGreen, xshift=3cm, yshift=2cm] (0, 0) rectangle (1,1);
      \node[above] at (2.5, 5) {$\hat{v}_{n}^B(\eta_b)$};
      \node at (-0.5,0) {};
      \node at (5.5,0) {};
    \end{tikzpicture}
    \cdot
    {\color{mRed}{e^{-F_n^\eta|\eta_a - \eta_b|}}}
  \]
  \caption[Schematic representation of Model B]{Schematic representation of Model B. Each point in $v_n(\eta)$ affects several elements in \hvnnee. The factor $e^{-F_n^\eta|\eta_a - \eta_b|}$ attenuates \hvnnee along $|\deta|$.}
  \label{fig:schema-model-b}
\end{figure}

The second model presented here was suggested by the CMS collaboration, albeit based on a vastly different analysis method~\cite{1503.01692}.
The model is given by
\begin{equation}
  \label{eq:cms-model}
  \hvnn(\eta_a, \eta_b) = \hat{v}_n^B(\eta_a) \hat{v}_n^B(\eta_b) e^{-F_n^\eta |\eta_a - \eta_b|}
\end{equation}
where the parameter $F_n^\eta$ is a measure for a $\Delta\eta = \eta_a - \eta_b$ dependent factorization breaking.
Despite being empirical $F_n^\eta$ provides insights into longitudinal fluctuations during the early stages of the collision~\cite{1011.3354,1511.04131}.
It should be noted that $\hat{v}_n^A(\eta) \neq \hat{v}_n^B(\eta)$ unless $F_n^\eta = 0$.
Analogous to the previous model, the flow coefficients $\hat{v}_n^B(\eta)$ and the constant $F_n^\eta$ are found by solving
\begin{equation}
  \label{eq:equation-sys-mod-b}
  V_{n\Delta}(\eta_a^i, \eta_b^j) = \hat{v}_n^B(\eta_a^i) \hat{v}_n^B(\eta_b^j) e^{-F_n^\eta |\eta_a^i - \eta_b^i|}
\end{equation}
The graphical representation of Eq.~\eqref{eq:equation-sys-mod-b} is depicted in Fig.~\ref{fig:schema-model-b}.
The exponential factor causes an attenuation of \hvnnee along $|\deta|$ and is constant along $\eta_a + \eta_b$.

Particle pairs with a small separation in $\Delta\eta$ are commonly excluded from flow analyses as this region of the phases space is known to exhibit large non-flow contributions.
Furthermore, an experimentally measured $V_{n\Delta}(\eta_a, \eta_b)$ may exhibit acceptance gaps if various detector systems are combined in order to maximize the $\eta$ coverage.
Therefore, the procedure to numerically solve the equation systems in Eq.~\eqref{eq:equation-sys-mod-a} and Eq.~\eqref{eq:equation-sys-mod-b} needs to be able to be performed on arbitrary subsets of the \eeplane.
A minimization of a weighted sum of squares fulfills this requirement.
The weighted sum $S$ is given by
\begin{equation}
  \label{eq:fact-least-square}
  S = \sum_{i,j=1}^{N_a^{\rm bin}, N_b^{{\rm bin}}} {
    \left (  \frac{V_{n\Delta}(\eta_a^i, \eta_b^j) - M(\eta_a^i, \eta_b^j)}{\sigma_{n\Delta}(\eta_a^i, \eta_b^j)} \right )
  }^2
\end{equation}
where $N_a^{\rm bin}$ ($N_b^{\rm bin}$) represents the number of bins in the $\eta_a$ ($\eta_b$) and $M$ represent either Model A or Model B as defined in Eq.~\eqref{eq:equation-sys-mod-a} or Eq.~\eqref{eq:equation-sys-mod-b} respectively.
The uncertainty associated with each point of $\hvnn(\eta_a^i, \eta_b^j)$ is given by $\sigma_{n\Delta}(\eta_a^i, \eta_b^j)$.

\subsection{Factorization ratio}
\label{sec:factorization-ratio-def}
The agreement of \hvnnee with either of the two models is assessed by means of a factorization ratio $f_n(\eta_a, \eta_b)$ defined by
\begin{equation}
  \label{eq:fact-ratio-def}
  f_n(\eta_a, \eta_b) = \frac{\hvnn(\eta_a, \eta_b)}{M(\eta_a, \eta_b)}
\end{equation}
where $M(\eta_a, \eta_b)$ represents either model with the parameters fitted to the observed \hvnnee.
Note that $f_n(\eta_a, \eta_b)$ can be computed for the entire \eeplane even if $S$ was only minimized for a subset of it.

\section{Results}
\begin{figure}
  \centering
  \includegraphics{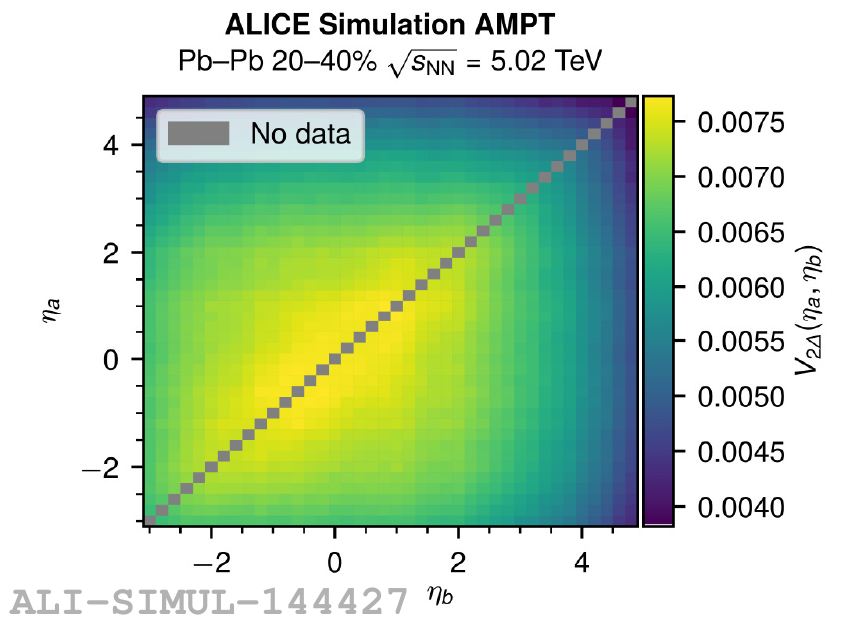}
  \caption[Factorization of \hvssee according to Model A]{
    The two-particle Fourier coefficients \hvssee{} from AMPT calculations of Pb--Pb collisions at \SI{5.02}{TeV} with a centrality of 20--40\%. The measurement of the two-particle Fourier coefficients is conducted for every point in the \eeplane independently.
  }\label{fig:V22}
\end{figure}

Figure~\ref{fig:V22} presents \hvssee obtained for collisions of 20--40\% centrality for AMPT calculations of Pb--Pb collisions at $\snn = \SI{5.02}{TeV}$ with string melting enabled.
Every two-particle Fourier coefficient in the \eeplane is computed independently with no a priori assumption about the event-by-event fluctuations.

\begin{figure}
  \centering
  \includegraphics[width=0.5\linewidth]{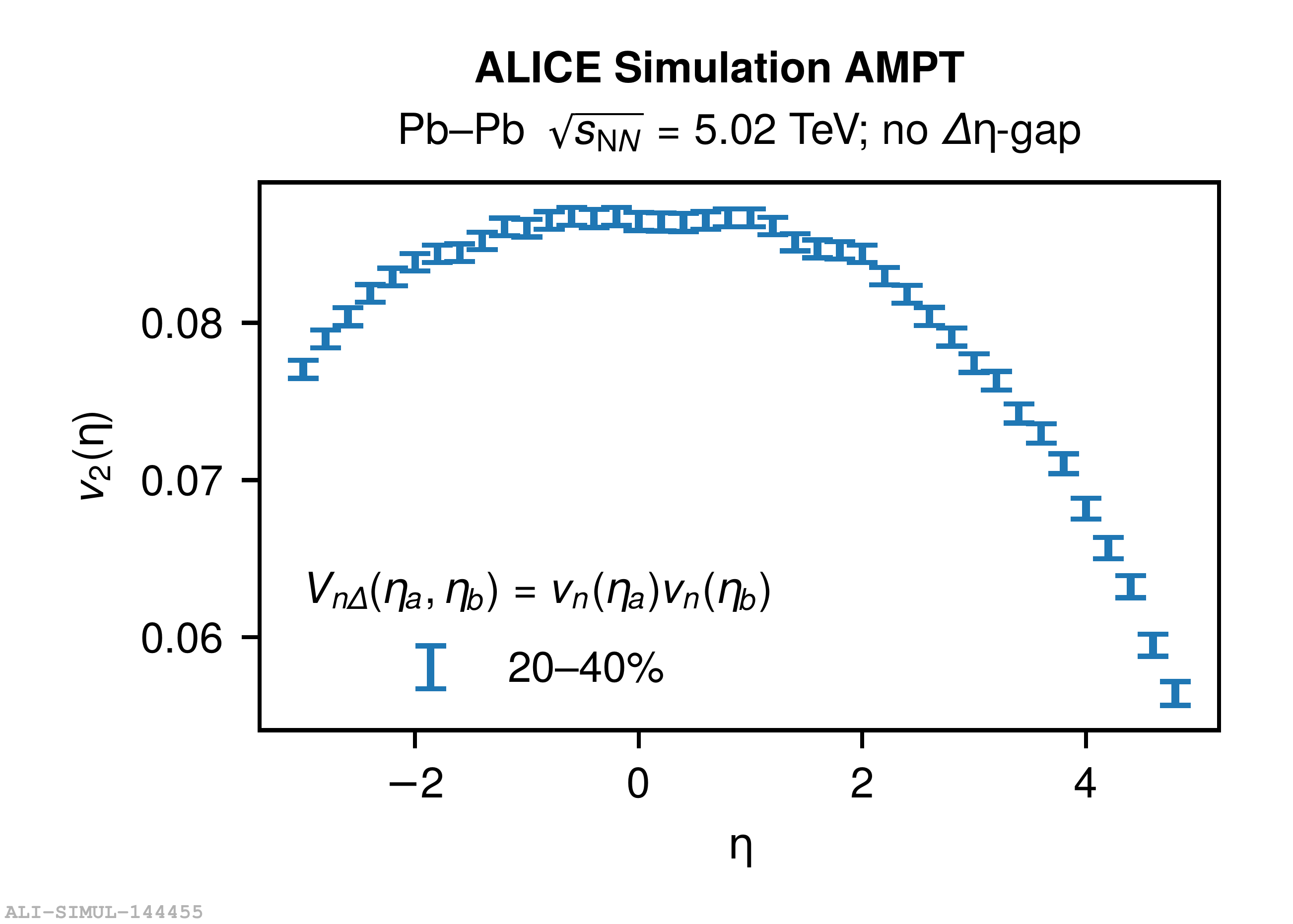}%
  \includegraphics[width=0.5\linewidth]{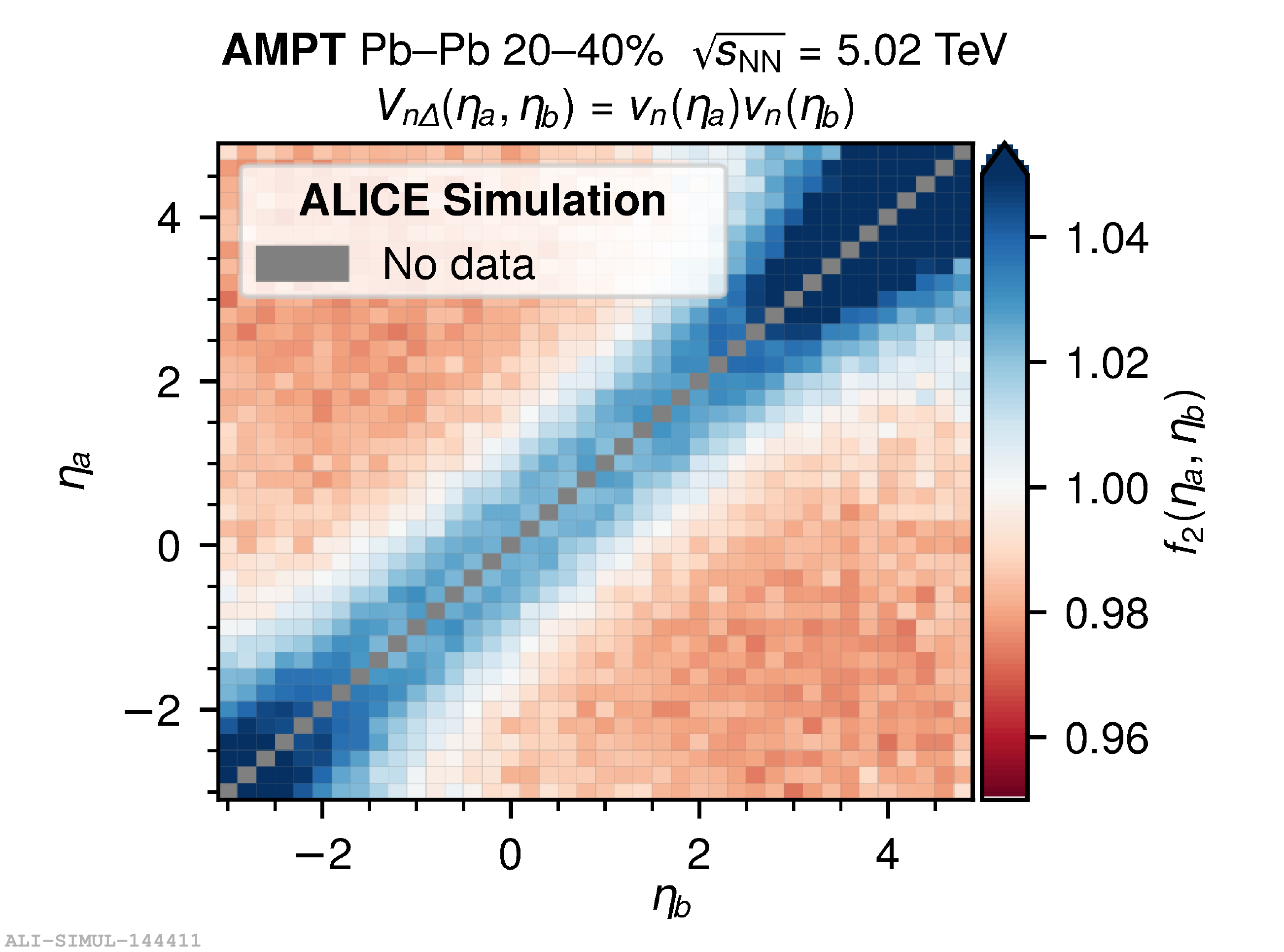}
  \caption[Factorization of \hvssee according to Model A]{
    Factorization of \hvssee according to Model A without the use of a \detagap based on AMPT simulations. Left: The flow coefficients yielded by the minimization procedure. Right: Factorization ratio for the solution shown on the left. The found solution fails to describe the short-range and long-range pairs.
  }\label{fig:v2_no_gap}
\end{figure}

Figure~\ref{fig:v2_no_gap} (left) presents $\hat{v}_n^A(\eta)$ obtained from a fit to \hvssee without the requirement of a \detagap.
Therefore, the factorization procedure included the short-range \deta region of the \eeplane which exhibits non-negligible non-flow contributions.
Figure~\ref{fig:v2_no_gap} (right) shows the factorization ratio for the flow coefficients from the left panel.
The short-range non-flow does not exhibit the same factorization behavior as the long-range regions.
This caused the fitted solution to neither accurately describe the short-range nor the long range regions of the \eeplane highlighting the need of a \detagap.

\begin{figure}
  \centering
  \includegraphics[width=0.5\linewidth]{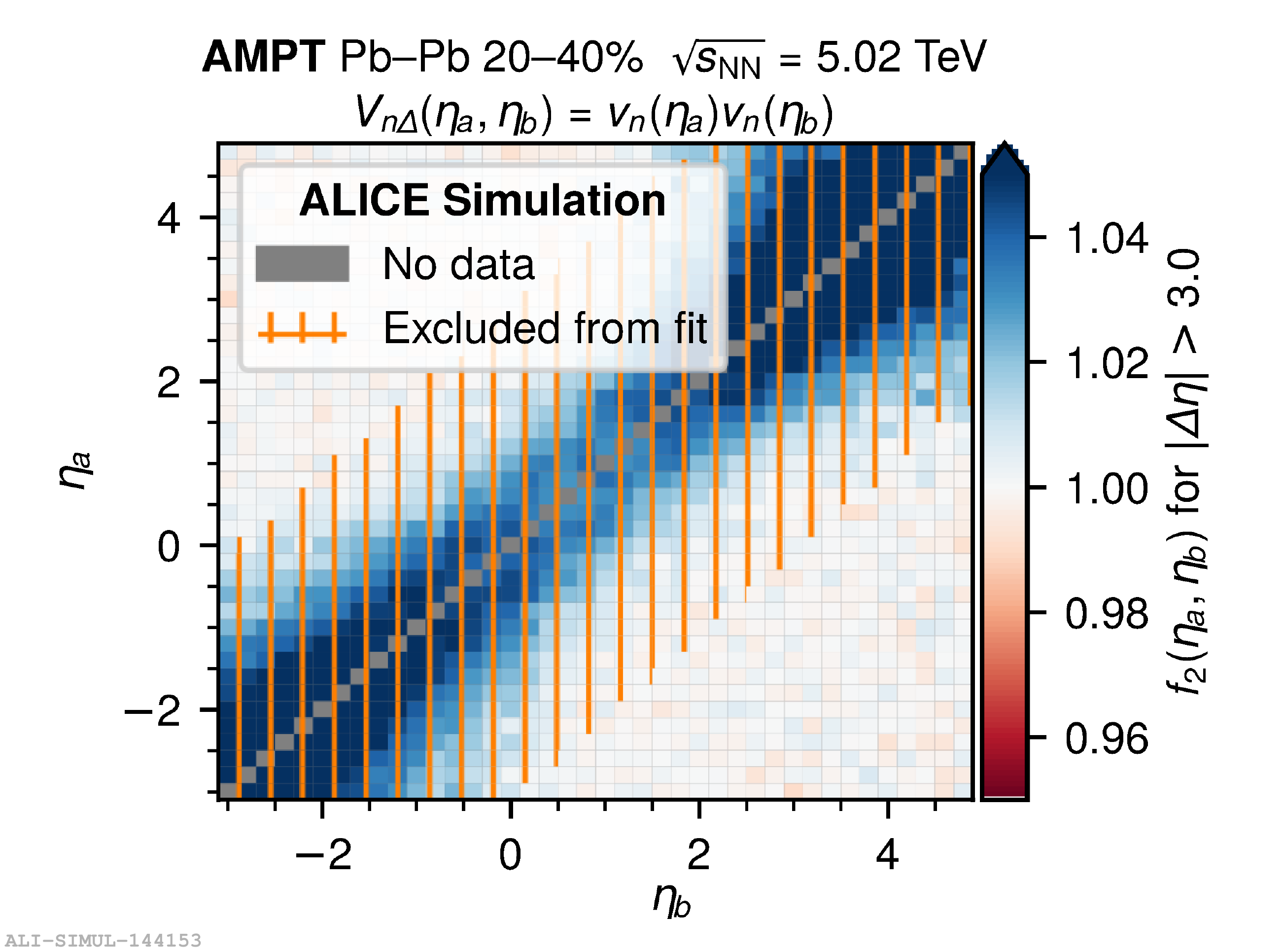}%
  \includegraphics[width=0.5\linewidth]{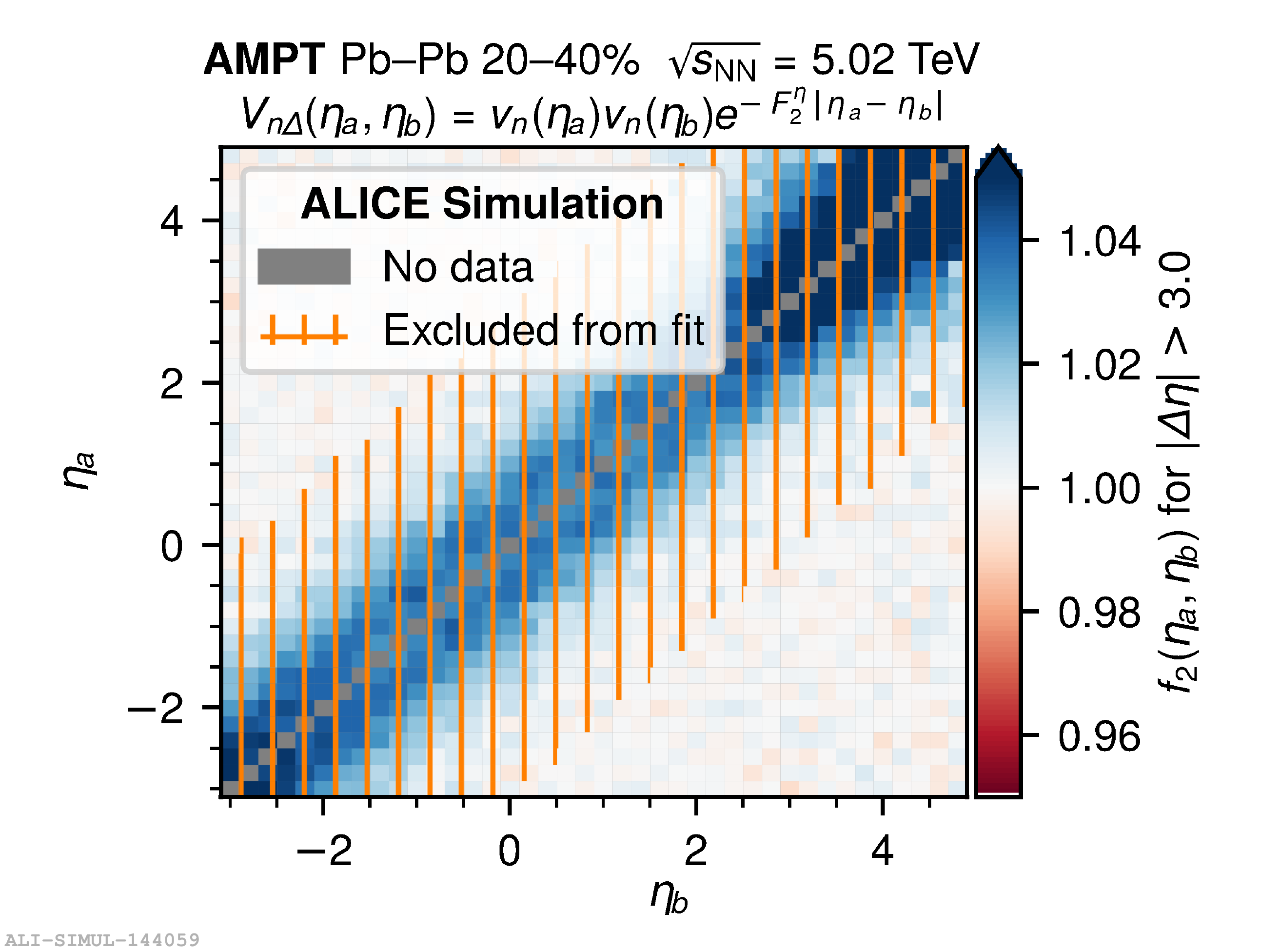}
  \caption{
    Factorization ratio for Model A (left) and Model B (right) with $|\deta| > 3$ in both cases.
    Either model provides good agreement for $|\deta|>3$, while the short-range region is better described by Model B.
  }\label{fig:fact_ratio_deta3}
\end{figure}

\begin{figure}
  \centering
  \includegraphics[width=0.8\linewidth]{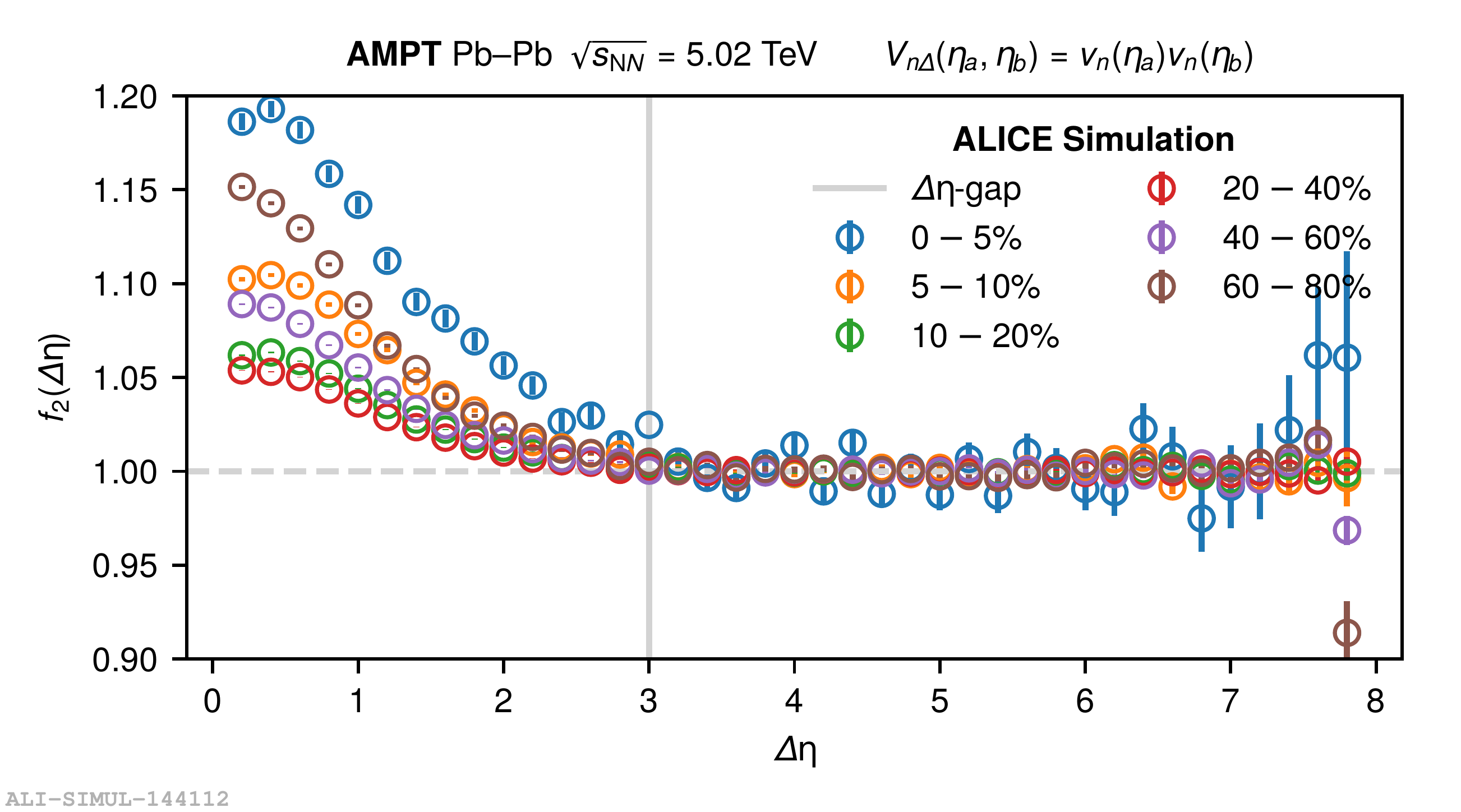}
  \caption{
    Projection of the factorization ratios of various centralities onto \deta.
    A \detagap of $3$ was imposed during the factorization procedure with Model A.
    The shown uncertainties are statistical.
  }\label{fig:deta_projection}
\end{figure}

Figure~\ref{fig:fact_ratio_deta3} (left) presents the factorization ratio for Model A if only particle pairs with $|\deta|>3$ are taken into account.
By excluding the short-range pairs, good agreement to Model A is observed in the long-range region.
However, the solution found for long-range pairs is not able to describe the short-range region of \hvssee further corroborating that non-flow effects in this region of the phase space do not factorize identically to the long-range anisotropic flow.
In order to determine the minimal \detagap necessary to exclude pairs originating from non-flow processes from the fitting procedure, a projection of the factorization ratios for Model A onto the \deta-axis is performed.
The results for all analyzed centralities and for a \detagap of $3$ are depicted in Fig.~\ref{fig:deta_projection}.
A centrality dependence for the factorization breaking in the short-range region is observed.
The deviation from unity is most pronounced for the most central events, decreases to a minimum for the 20--40\% centrality class and increases for more peripheral events thereafter.
This centrality dependence may originate from the centrality dependence of $\hat{v}_2^A(\eta)$.

Events of all centralities exhibit good agreement to Model A in the long-range region for $\deta > 3$.
A further increase of the \detagap does not significantly alter the extracted flow coefficients.
A decrease of the \detagap, includes short-range regions of the phase-space which are incompatible to the solution found in the long range region.
Factorization procedures with a smaller \detagap therefore decrease the fit quality in the long-range region.
Analyses which implicitly rely on the factorization assumption to hold should thus apply a minimal longitudinal separation of $\demin \approx 3$ for the kinematic region here studied.
The method presented here can be used to improve the precision of similar previously published results which found that the factorization assumption holds to within $10\%$ for Pb--Pb collisions at $\snn = \SI{2.76}{TeV}$ for $\deta > 2$~\cite{1203.3087}.

\begin{figure}
  \centering
  \includegraphics{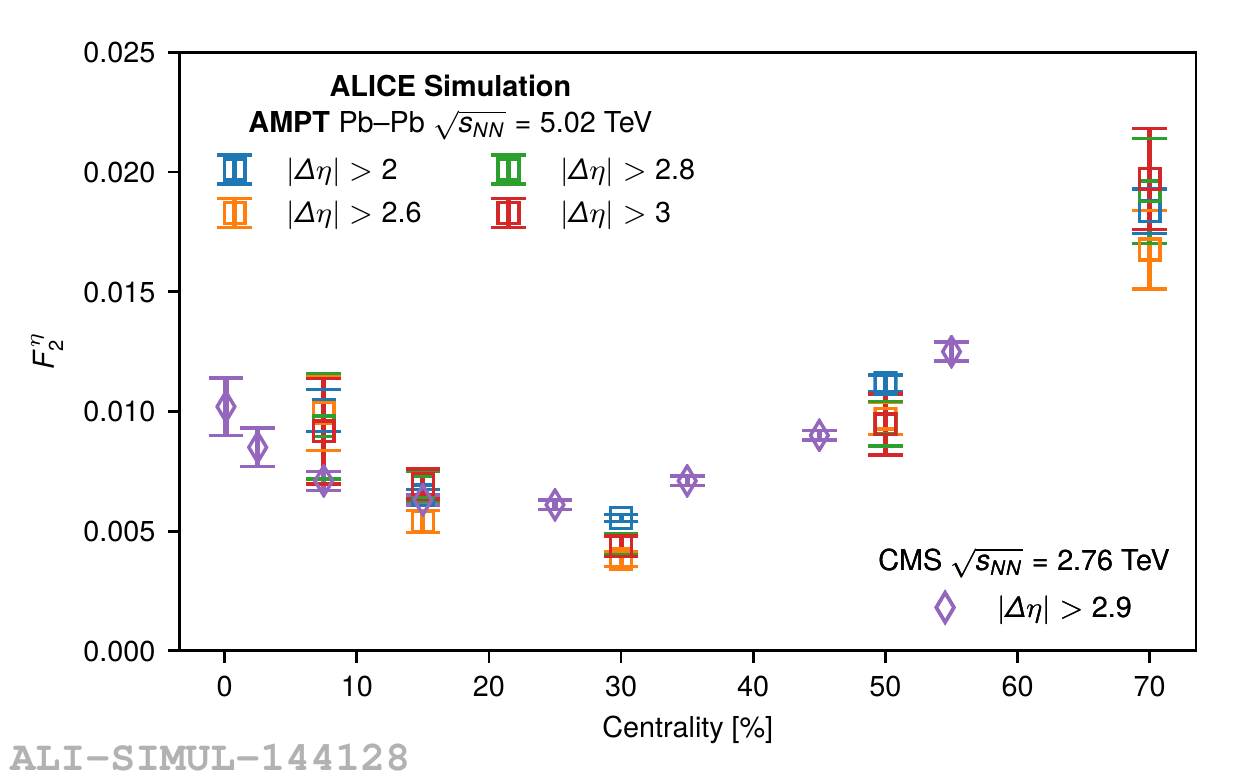}
  \caption{Empirical decorrelation parameter $F_2^\eta$ for AMPT model calculations of Pb--Pb collisions at $\snn = \SI{5.02}{TeV}$. The simulation is compared to measurements by the CMS collaboration for $\snn = \SI{2.76}{TeV}$.}
  \label{fig:F2}
\end{figure}

Performing the factorization using Model B allows for the measurement of the decorrelation parameter $F_2^\eta$.
The decorrelation parameters computed with the method presented here is shown in Fig.~\ref{fig:F2} for various \detagap{}s and centralities.
The data point for the 0--5\% centrality bin is removed from the figure due to a poor statistical uncertainties.
Results published by the CMS collaboration for Pb--Pb collisions at $\snn = \SI{2.76}{TeV}$ are included for comparison.
The analysis used by the CMS collaboration corresponds to a \detagap of approximately $2.9$ in this analysis.
The centrality dependence observed for Pb--Pb collisions is reproduced by the AMPT simulations at \SI{5.02}{TeV} supporting previous model comparisons~\cite{1511.04131} and studies of the energy dependence of $F_2^\eta$~\cite{1709.02301}.
The centrality dependence of the decorrelation parameter is also found to reflect the centrality dependence of the short-range factorization breaking in Fig.~\ref{fig:fact_ratio_deta3}.
However, a quantification of possible non-flow contributions to the centrality dependence of $F_2^\eta$ requires further research.

\section{Summary}
Depending on the nature of the event-by-event fluctuations, the two-particle Fourier coefficients \hvnnee may assume different functional shapes.
In these proceedings two distinct models were studied: The first model is a purely factorizing one as it is implicitly assumed in most flow analyses.
The second model is an extension to the former and allows for a \deta dependent attenuation of \hvnnee.
AMPT calculations of Pb--Pb collisions at $\snn = \SI{5.02}{TeV}$ were used as the basis for the  presented results.
The first model was use to estimate the longitudinal extent of short-range non-flow correlations under the assumption that such effects are not well described by the factorized solution found from long-range particle pairs.
For the studied kinematic region a minimal \deta-separation of $\demin \approx 3$ is required for the factorization assumption to hold in the long-range region.

The second model was used to determine \deta-dependent decorrelation effects as they are to be expected from event-plane decorrelation effects.
The empirical decorrelation parameter $F_2^\eta$ is qualitatively compatible to measurements by the CMS collaboration at $\snn = \SI{2.76}{TeV}$.
This confirms previous studies suggesting that AMPT is able to reproduce the observed decorrelation effects as well as that these effects exhibit only a weak dependence on the center of mass energy~\cite{1709.02301,1511.04131}.

The method presented here offers a new way to investigate possible non-flow contributions to the observed decorrelation effects and will help to better understand the three dimensional initial conditions of heavy ion collisions in the future.

\bibliography{sources}

\end{document}